# Infrared uplink design for visible light communication (VLC) systems with beam steering

Osama Zwaid Alsulami
*School of Electronic and Electrical Engineering*
*University of Leeds*
Leeds, United Kingdom
ml15ozma@leeds.ac.uk

Mohammed T. Alresheedi
*Department of Electrical Engineering*
*King Saud University*
Riyadh, Saudi Arabia
malresheedi@ksu.edu.sa

Jaafar M. H. Elmirghani
*School of Electronic and Electrical Engineering*
*University of Leeds*
Leeds, United Kingdom
j.m.h.elmirghani@leeds.ac.uk

*Abstract*— Providing uplink high data rate is one of the big concerns in visible light communication (VLC) systems. This paper introduces an uplink VLC system based on an infrared transmitter with beam steering to provide high data rates. In this work, a 4 branches angle diversity receiver (ADR) is used and the resultant delay spread and SNR are examined. The proposed system achieved data rates up to 3.57 Gb/s using simple on-off-keying (OOK).

*Keywords*— VLC, IR, beam steering, angle diversity receiver, SNR, data rate.

## I. INTRODUCTION

Due to the growth in demand for high data rates, visible light communication (VLC) has received increasing interest by researchers. The radio frequency (RF) spectrum is currently the main medium for wireless communication. However, traditional radio communication systems suffer from limited channel capacity and transmission rate due to the limited radio spectrum available, while the data rates requested by the users continue to increase exponentially. Achieving very high data rates (beyond 10 Gbps and into the Tbps regime) using the bandwidth of radio systems is challenging. According to Cisco, mobile Internet traffic over this half of the decade (2016-2021) is expected to increase by 27 times [1]. Given this expectation of dramatically growing demand for data rates, the quest is already underway for alternative spectrum bands beyond radio waves. The latter are bands currently used and planned for near future systems, such as 5G cellular systems [2]. Optical wireless (OW) systems and VLC systems can provide license free bandwidth, high security and low-cost compared to RF system [3]–[10]. However, VLC systems have some limitations and one of these limitations is the absence of the line-of-sight (LOS) components in the link which reduces the system's performance significantly [11], [12]. In addition, inter-symbol interference (ISI), which is caused by multi-path propagation in VLC systems, can affect the system's performance as well. Moreover, uplink communication is one of the biggest challenges in VLC systems. Using VLC as an uplink results in high glare affecting the users' eyes. As a result, VLC is suited for downlink communication and infrared (IR) can be used for the uplink communication.

VLC systems have been shown in many studies to be able to transmit video, data and voice, at data rates up to 20 Gbps in indoor systems [13]–[17]. However, it provides only one direction of communication, i.e. downlink. Therefore, it is challenging to provide bidirectional (downlink and uplink) communication in a VLC system unless another technology such as RF or IR is used for the uplink communication. Many different techniques have been investigated for providing uplink communication in VLC systems. In [18], the authors proposed the use of retro reflecting transceivers along with a corner cube modulator to offer a low-speed uplink connection. Another technique involves the use of hybrid systems such as an RF system with a VLC system. This technique uses the RF system for uplink communication and the VLC system for downlink communication [19]. A bidirectional transmission technique was discussed in [20]. It uses a VLC system for both uplink and downlink communication, resulting in a high data rate of 575 Mbps for the downlink communication and 225 Mbps for the uplink communication. This technique uses subcarrier modulation (SCM) combined with wavelength division multiplexing (WDM). In addition, an uplink was proposed recently for VLC systems using a beam steered infrared communication (IRC) setup. It provided a data rate up to 2.5 Gbps [21].

In this paper, uplink communication in a VLC system is introduced. IR is used in the transmitter to provide uplink communication and unlike [21], an ADR is utilised as an optical receiver to reduce the effects of ISI. This results in enhancing the performance of the system beyond the performance of [21]. The effects of mobility and multi-path propagation are considered in this paper. This paper is organised as follows: The room configuration is described in Section II. The optical receiver design is given in Section III. Section IV illustrates the optical transmitter design. Section V illustrates the simulation results and conclusion are drawn in Section VI.

## II. ROOM CONFIGURATION

As shown in Fig. 1 the room dimensions (length × width × height) are assumed to be 8 m × 4 m × 3 m in the simulation. An empty room was considered in the simulation with no doors or windows. Reflections up to second order were considered in this work since the third and higher order reflections have negligible impact on the received optical power [13], [22]. A ray-tracing algorithm was utilised to model reflections from the ceiling, walls and the floor in the room. Thus, each surface in the room was divided into small equal areas of size $dA$ with a reflection coefficient of $\rho$. The authors in [23] showed that plaster walls reflect light rays close to a Lambertian pattern. Therefore, each surface in the room such as ceiling, walls and floor was modelled as a Lambertian reflector with reflection coefficient equal to 0.8 for ceiling and walls and 0.3 for the floor. Each element in each surface acts as a small emitter that reflects the received ray in the form of the Lambertian pattern with $n$ (emission order of the Lambertain pattren) equal to 1. The area of the surface elements can play a significant impact in the resolution of the results. When the surface elements is very small a higher resolution result can be obtained but that comes at the cost of long computation time in the simulation. Therefore, 5 cm × 5 cm was chosen as an area of the surface element for the first order reflection, while 20 cm × 20 cm was chosen as an area of the surface element for the second order reflection to keeping the computation time of the simulation within a reasonable time [15], [24]. The communication floor (CF) is set as 1 m above the floor as shown in Fig. 1 which means that all communications are performed on the CF.

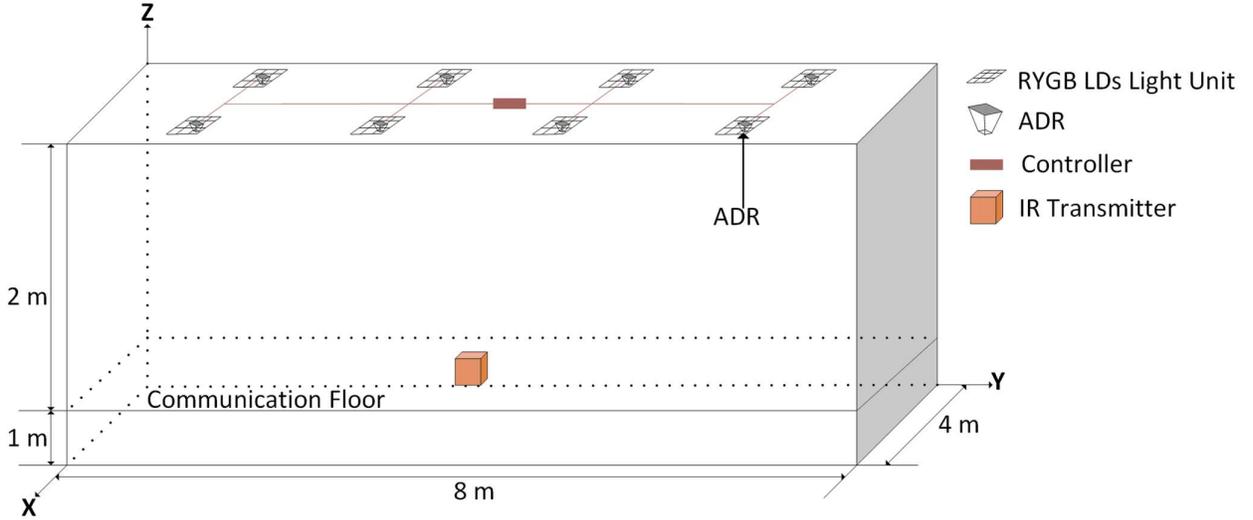

Figure 1. Room Configuration.

### III. OPTICAL RECEIVER DESIGN

In this paper, a 4 face ADR (see Fig. 2) with narrow fields of view (FOVs) is used to collect signals and reduce the inter-symbol interference (ISI). These units are placed on the ceiling in different locations at (1 m, 1 m, 3 m), (1 m, 3 m, 3 m), (1 m, 5 m, 3 m), (1 m, 7 m, 3 m), (3 m, 1 m, 3 m), (3 m, 3 m, 3 m), (3 m, 5 m, 3 m) and (3 m, 7 m, 3 m). Each unit has 4 branches with photodetectors as shown in Fig. 2 that cover a specific area of the room, 2 m × 2 m, and each photodetector is oriented to a different direction to cover a small different area in the room based on two angles: Azimuth (Az) and Elevation (El). The *El* angles of the four detectors are set at -70°. The *Az* angles of detectors are 45°, 135°, 225° and 315°. The FOV of these detectors are set at 21°. In addition, the area of each photodetector was chosen equal to 4 mm$^2$ with responsivity equal to 0.4 A/W.

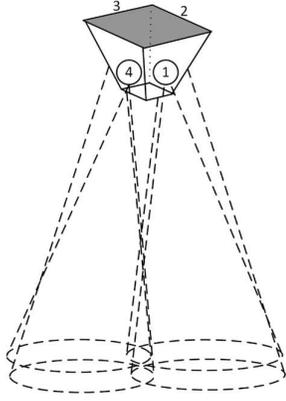

Figure 2. Optical Receiver Design.

### IV. OPTICAL TRANSMITTER DESIGN

In this paper, an infrared transmitter was used for uplink communication. The semi-angle of the transmitter was set at 40⁰ to see at least one receiver unit at any location in the room. The transmit power was chosen to be small to cater for the eye safety requirements in the indoor environment and is set equal to 150 mW. The transmitter was examined along x = 2 and different y-axis locations: y = 1 m, 2 m, 3 m, 4 m, 5 m, 6 m and 7 m. An optical hologram similar to [25] was used in front of the transmitter to provide beam steering with the ADR [3], [26], [27], (with possible extension to imaging receivers [28]). As a result, the received power increased.

The hologram is used to increase the SNR and reduce the ISI by focusing the beam on the receiver location using beam steering. The adaptive hologram implemented using a liquid crystal device works as follows:

- The coverage area of the Infrared transmitter is divided into four quadrants based on its transmission angles, which are from -40⁰ to 40⁰, to choose the best quadrant that has a highest SNR to become a new coverage area.

- Then, the new coverage area is divided into four sub-quadrants and this step is repeated until the receiver location is identified.

- Subsequently, the beam is steered to the receiver location.

### V. SIMULATION SETUP AND RESULTS

In this work, the simulation starts by dividing the transmitter coverage area into 4 quadrants in order to choose the best ADR branch based on the SNR. Then, the transmitter sends a setup signal to each quadrant starting by the first one and waits for feedback from the controller. At the receiver, the SNR is calculated and transferred to the controller through a fibre optic connection with information about the angles of the hologram used in the transmitted signal. Then, feedback is sent to the transmitter with the angles that provide a higher SNR. After that, the chosen quadrant is divided into 4 sub-quadrants to identify the sub-optimum location. This process is repeated until the sub-quadrant area becomes 10 cm × 10 cm. The entire transmitter power is then oriented to the receiver location using beam steering to provide high SNR. The proposed system was compared with a similar system but without using beam steering. The proposed system reduced the delay spread significantly from 0.01 ns to 0.0001 ns when the transmitter is placed at the room centre (2 m, 4 m, 1 m) as shown in Fig. 3a. In addition, the SNR was calculated based on the best branch of the ADR and the proposed system provides a significant improvement in SNR compared to the normal system without beam steering (at high data rates up to 3.57 Gb/s) as shown in Fig. 3b.

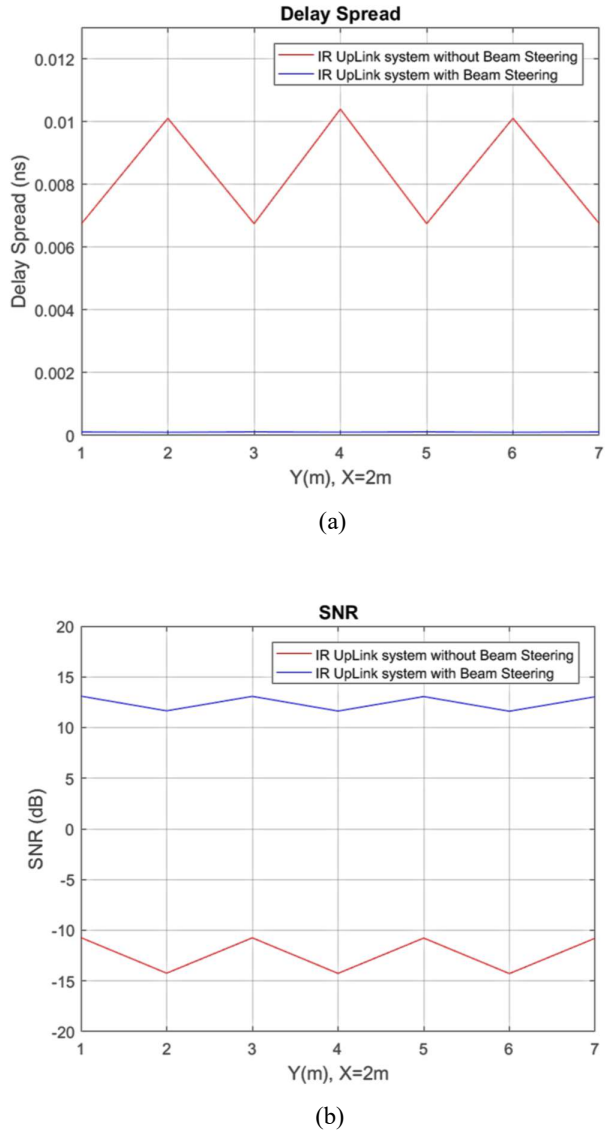

Figure 3. (a) Delay Spread, (b) SNR.

## VI. CONCLUSIONS

An IR system was proposed in this paper for uplink communication in VLC systems. It can offer a high data rate, up to 3.57 Gb/s, over the entire room while using simple OOK modulation. In addition, a 4 branch ADR was used to reduce the delay spread and improve the SNR. The proposed system provides a high improvement in the delay spread and SNR at high data rate up to 3.57 Gb/s.

## ACKNOWLEDGMENT

The authors would like to acknowledge funding from the Engineering and Physical Sciences Research Council (EPSRC) for the TOWS project (EP/S016570/1). All data are provided in full in the results section of this paper.